# Payload-size and Deadline-aware Scheduling for Upcoming 5G Networks: Experimental Validation in High-load Scenarios

Stefan Monhof, Marcus Haferkamp, Benjamin Sliwa, Christian Wietfeld
Communication Networks Institute (CNI)
TU Dortmund University, 44227 Dortmund, Germany
Email: {Stefan.Monhof, Marcus.Haferkamp, Benjamin.Sliwa, Christian.Wietfeld}@tu-dortmund.de

*Abstract*—High data rates, low latencies, and a widespread availability are the key properties why current cellular network technologies are used for many different applications. However, the coexistence of different data traffic types in the same 4G/5G-based public mobile network results in a significant growth of interfering data traffic competing for transmission. Particularly in the context of time-critical and highly dynamic Cyber Physical Systems (CPS) and Vehicle-to-Everything (V2X) applications, the compliance with deadlines and therefore the efficient allocation of scarce mobile radio resources is of high importance. Hence, scheduling solutions are required offering a good trade-off between the compliance with deadlines and a spectrum-efficient allocation of resources in mobile networks. In this paper, we present the results of an experimental validation of the Payload-size and Deadline-aware (PayDA) scheduling algorithm using a Software-Defined Radio (SDR)-based eNodeB. The results of the experimental validation prove the high efficiency of the proposed PayDA scheduling algorithm for time-critical applications in both miscellaneous and homogeneous data traffic scenarios.

## I. INTRODUCTION

Considering the increasing diversity of applications with various requirements regarding the data transmission, especially massive Machine-Type Communication (MTC) in 5G, the efficient allocation of confined radio resources is a key component of mobile networks and will become even more important for real-time Vehicle-to-Everything (V2X) and Cyber Physical Systems (CPS) applications in the future. As a result, there is a negative impact on the compliance with Quality of Service (QoS) requirements. Especially in the context of highly dynamic and time-critical applications such as Cooperative Awareness Message (CAM), the compliance with deadlines is one of the most important criteria. In contrast to commonly used real-time scheduling techniques, the Payload-size and Deadline-aware (PayDA) scheduling mechanism, proposed in [1], takes deadlines as well as payload-sizes into account, allowing a high-efficient transmission of time-critical data in miscellaneous high-load data traffic scenarios. Fig. 1 exemplifies the prioritization of User Equipments (UEs) in a simple scenario due to deadlines regarding the data transmission and the current size of the Medium Access Control (MAC) queue. While our previous work focused on the simulative evaluation of PayDA [1], in this paper, we provide the results of an experimental validation based on extensive measurements in the laboratory with multiple UEs and a Software-Defined Radio (SDR)-based Evolved Node B (eNodeB). All measurement results are provided online as open access data [2]. The simulation-based implementation of PayDA is published open source [3].

The rest of this paper is structured as follows: The related work is discussed in Sec. II, which is further followed by the presentation of the PayDA scheduling strategy in Sec. III. In Sec. IV, we present the implementation of PayDA for an SDR-based eNodeB, the experimental setup used for validation, and detailed results on various Key Performance Indicators (KPIs) of data transmission (e.g., Deadline-Miss-Ratio (DMR) and latency) for different data traffic patterns in Long Term Evolution (LTE). The results presented and discussed in Sec. V show that PayDA can be used in various scenarios due to its good scaling properties for an increasing number of users in a cell and its drastic reduction of the average DMRs.

## II. RELATED WORK

Different resource scheduling mechanisms that focus on specific goals like spectral-efficient resource allocation or enabling highly time-critical applications are used in mobile networks. A detailed overview of common scheduling strate-gies used for LTE downlink transmissions is given in [4], including an individual assessment of the suitability of these procedures for different kinds of Human-to-Human (H2H) data traffic. Taking the need for QoS support into account, there are different approaches for two-layered scheduling mechanisms, which statically classify data traffic as Real-Time (RT) or Non-Real-Time (NRT) data streams. The approach presented in [5] is analyzed for high-rate and time-critical H2H multimedia data traffic, exposing a high suitability for real-time data traffic due to low DMRs, but at the expanse of significant low mean data rates in case of best effort data traffic. In contrast, the two-layered scheduling approach proposed in [6] is evaluated for smart grid and H2H data traffic performing the resource allocation according to each user's Packet Loss Ratio (PLR). Nonetheless, the static and binary classification of data traffic as RT and NRT does not seem to be well-suited for a voluminous and equal growth of both data traffic types with manifold QoS requirements. For this reason, the usage of universal scheduling strategies respecting each bearer's data traffic individually is a reasonable idea. Such a universal scheduling mechanism is proposed in [7] for LTE downlink transmissions combining the core features of other scheduling strategies respecting each buffer's queue size and Head-of-Line (HOL) delay, the maximum probability that the HOL packet exceeds its deadline, and the ratio of channel quality and achieved average data rate. Considering MTC uplink data traffic scenarios, different approaches are presented in [8] and [9] posing to be spectral-efficient or buffer-aware real-time scheduling strategies, respectively. The use of different application types in the context of mobile networks causes interdependences of different data traffic types, which is evaluated for time-critical H2H and time-tolerant MTC data traffic in [10]. However, the previously mentioned scheduling approaches are only evaluated for typical H2H or MTC data traffic, respectively. In our previous work [1], we have evaluated the proposed PayDA scheduling strategy for manifold data traffic respecting the interdependence of different RT and NRT traffic flows by simulations. In this paper, we validate the high suitability of PayDA for different data traffic scenarios by extensive measurements in the laboratory.

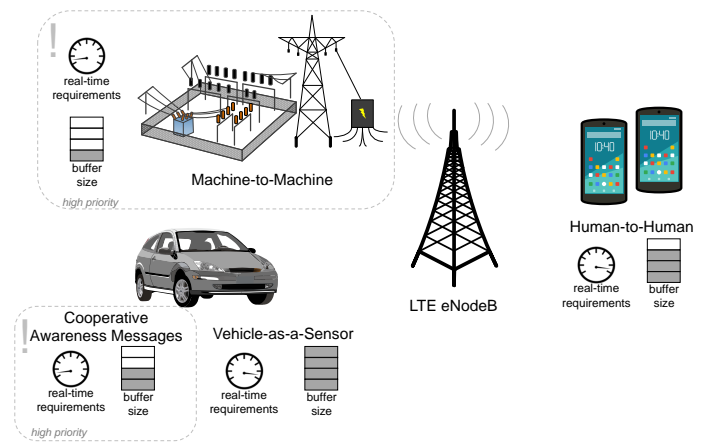

Figure 1. Example scenario for prioritization of UEs concerning the allocation of spectral radio resources based on temporal requirements regarding the data transmission, and the current buffer size. PayDA schedules application data that requires low delays and is small-sized with a higher priority.

## III. PAYLOAD-SIZE AND DEADLINE-AWARE SCHEDULING

The PayDA scheduling approach is a RT-capable, resource-efficient and low-complexity packet scheduling strategy for current and future mobile networks based on the common RT-aware scheduling procedure Earliest Deadline First (EDF). In contrast to EDF, PayDA is enhanced by additionally considering the remaining packet size of each data flow. The main idea of EDF is to schedule the packet with the closest deadline for data transmission, which is achieved by evaluating Eq. 1 for each user:

$$\omega_i = \max\left(0, \frac{1}{(\tau_i - D_{HOL,i})}\right) \quad (1)$$

where $\omega_i$ is the calculated metric used for resource allocation, $\tau_i$ is the deadline and $D_{HOL,i}$ is the HOL delay of the first packet in the buffer of the *i*-th user. Consequently, the HOL delay $D_{HOL}$ marks the time period a packet has been queued since its generation and thus can be used for calculating the remaining time to its specific deadline ($\tau_i - D_{HOL,i}$). The less time up to a certain deadline is left, the higher the resulting scheduling metric $\omega_i$ for the data flow of the *i*-th user is. By selecting the maximum of 0 and the calculated metric, negative metric values are prevented, which possibly cause starvation of UEs in high-load scenarios.

However, the EDF scheduling procedure only aims to achieve a compliance with deadlines without considering further essential aspects in the context of mobile networks. Therefore, the metric calculation of Eq. 1 was enhanced by additionally respecting each user's remaining HOL packet size:

$$\omega_i = \max\left(0, \frac{1}{((\tau_i - D_{HOL,i}) \cdot \delta_{left,i})}\right) \quad (2)$$

where $\delta_{left,i}$ is the remaining HOL packet size of the *i*-th user. Apparently, both aspects ($\tau_i - D_{HOL,i}$) and $\delta_{left,i}$ are of equal importance in terms of the metric calculation. Therefore, the PayDA scheduling strategy tends to prefer time-critical and low-volume data traffic compared to either time-tolerant or high-volume data traffic types. In this way, PayDA enables an efficient packet scheduling with respect to RT requirements.

## IV. EXPERIMENTAL PERFORMANCE VALIDATION

In this section, the laboratory-based system setup that is used for the experimental performance validation of PayDA is presented. In this regard, relevant aspects of the software-based implementation as well as further details of the experimental laboratory setup and used data traffic classes are given.

### A. SDR-based Implementation

To validate the performance of the PayDA scheduling algorithm, it was implemented as a module for the *CommAgility SmallCellSTACK*, which is a 3rd Generation Partnership Project (3GPP) Rel. 9 compatible eNodeB implementation for SDR platforms. The task of PayDA is to calculate the scheduling metric for each Dedicated Radio Bearer (DRB) separately, resulting in the support of multiple DRBs per

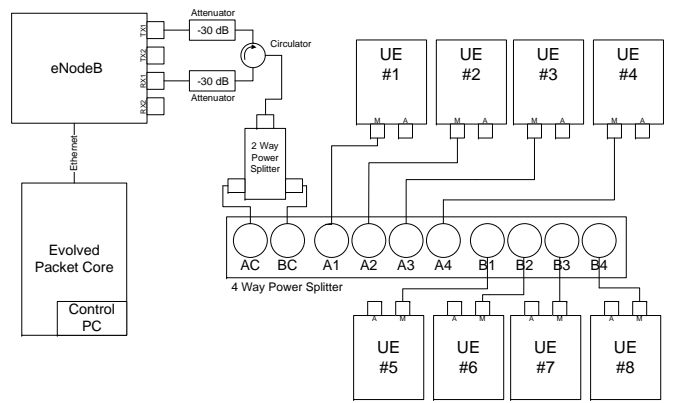

Figure 2. Lab setup for experimental performance validation of PayDA and further common scheduling algorithms.

Table I
PARAMETERS USED FOR THE EXPERIMENTAL VALIDATION OF PAYDA IN THE LABORATORY SETUP AND SIMULATIONS.

| Parameter | Value |
| --- | --- |
| System bandwidth | 5 MHz (25 Resource Blocks) |
| Operating frequency | 2.6 GHz |
| RLC mode | Unacknowledged Mode (UM) |
| Number of User Equipments | 1 to 30 (simulation) / 1 to 8 (lab) |
| Total measurement period | 300 s (simulation) / 100 s (lab) |
| Start times of applications | Equally distributed in the first second |
| Simulation/measurement runs | 10 (simulation) / 5 (lab) |

UE. For the actual allocation of Resource Blocks (RBs), the individual DRBs of each UE are multiplexed and scheduled by the corresponding scheduling metric. At this point, the scheduled DRB with the highest scheduling metric receives the needed amount of available RBs. In order to analyze the performance of the PayDA scheduling algorithm, essential KPIs of each data transmission, in particular each UE's HOL delay, DMR and data rates, are extracted directly from the Radio Link Control (RLC) and scheduler modules in the eNodeB. In this case, the scheduler calculates the moving average of relevant KPIs for a better comparability on a uniform time basis. As traffic generator we use a custom build application written in C, enabling a precise adjustment of the data traffic pattern including the amount and frequency of data transmissions (e.g., data size and transmission interval). Finally, we chose User Datagram Protocol (UDP) as transport protocol for our performance analysis to prevent interfering side effects such as the adaption of data rates by congestion control algorithms.

*B. Laboratory Setup*

The experimental validation of PayDA is performed by measurements in a wired laboratory setup. In this setup, a *CommAgility AMC-K2L-RF2* SDR platform running *CommAgility SmallCellSTACK* is used as eNodeB, allocating RBs to custom-built UEs based on embedded PCs and *Huawei ME909s-120* LTE modules. As shown in Fig. 2, eNodeB and UEs are connected using Radio Frequency (RF) combiners and a circulator to split and combine downlink and uplink signals. Due to the wired setup, all UEs have comparable channel conditions with a downlink Channel Quality Indicator (CQI) of 15, which results in a high spectral efficiency. Since PayDA does not take channel conditions into account, this fact is not crucial for the experimental performance validation. The eNodeB is connected to the Evolved Packet Core (EPC), which also contains a control PC that is on one hand used for generating the downlink traffic required for the performance validation and on the other hand for monitoring and logging of relevant KPIs. In Tab. I, the most important parameters for both laboratory setup and simulations are given. Specific details of the used data traffic characteristics are presented in the following subsection.

*C. Data Traffic Classes*

In order to quantify the performance of PayDA for realistic scenarios, different data traffic patterns on the basis of various data traffic classes were analyzed by the experimental validation and simulations. Tab. II lists QoS requirements of typical consumer and business applications such as file transfer, web surfing or Voice-over-IP (VoIP), which are less critical in terms of sporadic or minor delays. In addition, time-critical applications like voltage control and power flow optimization in the field of smart grids [12] or automatic train control [13] are given. Each application listed in Tab. II is assigned to at least one appropriate LTE QoS Class Identifier (QCI) [14]. These applications with different requirements and characteristics are the basis for realistic data traffic scenarios that were used to validate PayDA and further common scheduling mechanisms. With regard to the performance analysis for this paper, each UE runs only one application and receives the corresponding data traffic. This approach enables a UE-specific evaluation of the effects on data traffic characteristics by using different schedulers. Regarding the experimental laboratory setup, the main challenge is to define realistic and high-load data traffic scenarios for a comparatively small number of available UEs that also scale for real-world scenarios with significantly more users. Those high-load data traffic scenarios are necessary for a high cell utilization to evaluate the effects of various scheduling strategies on the resulting data traffic characteristics. Subsequently, the scenarios chosen for the experimental validation of different schedulers are discussed in detail. Obviously, these scenarios are only a subset of data traffic scenarios and can be enhanced by adding further application types.

In the first scenario, the performance results of PayDA and further common schedulers achieved by simulations and the lab-based measurements are compared and validated for *homogeneous data traffic*. In this case, the overall data traffic consists of data packets of equal size, which are sent in constant intervals to up to eight UEs. In this scenario, the only distinctive feature for RT and NRT data traffic is the time-sensitivity in terms of the delay margin, which is of high importance in case of RT schedulers such as PayDA. Here, one UE is defined as RT user with a delay margin of $100\,\text{ms}$ (QCI 7), whereas the remaining users act as NRT users with a three times larger delay margin of $300\,\text{ms}$ (QCI 9). The traffic pattern of the used RT and NRT applications corresponds to those of the *crash avoidance* application as listed in Tab. II, with a transmission interval of $100\,\text{ms}$ and a data rate of $2666\,\text{kbit/s}$. Since the maximum data rate with $5\,\text{MHz}$ cell bandwidth and Single Input Single Output (SISO) mode is approximately $18.3\,\text{Mbit/s}$, a full utilization of the downlink capacity can be achieved with a total of seven users.

In the second scenario, the focus is on *heterogeneous data traffic* generated by various application types differentiating in terms of data size and transmission interval. In total, three RT applications are used: a voltage control application (QCI 5) with a data rate of $625\,\text{kbit/s}$ and a transmission interval of $1000\,\text{ms}$, a crash avoidance application (QCI 7) with a data rate of $500\,\text{kbit/s}$ and an interval of $50\,\text{ms}$, and the crash avoidance application (QCI 9) from the homogeneous scenario ($2666\,\text{kbit/s}$ and an interval of $100\,\text{ms}$). The remaining NRT

Table II
APPLICATION PARAMETRIZATION USED FOR VALIDATION OF PAYDA

| Application | Max. delay [s] | Payload [Byte] | Interval [s] | Data rate [kbit/s] | LTE QCI |
|---|---|---|---|---|---|
| VoIP | 0.1 | 32 | 0.02 | 12.8 | 1/3 (GBR), 5/7 (Non-GBR) |
| Video (H.264) | 1 | variable | 0.4 | 242 | arbitrary |
| Video call (High Quality / High Definition) | 0.2 | variable | - | 500 / 1,500 | 1/2/3 (GBR), 5/7 (Non-GBR) |
| Website | 10 | $5 \cdot 10^3$ | - | - | arbitrary |
| File Transfer | 60 | $5.25 \cdot 10^6$ | 30 | - | arbitrary |
| Voltage control and power flows optimization | 0.025 - 0.1 | variable | 0.5 - 5 | 2,000 - 5,000 | 1/3 (GBR), 5/7 (Non-GBR) |
| Crash avoidance (smart road/train control) | 0.1 | variable | 0.01 - 0.1 | 2,000 - 6,000 (peak) | 1/3 (GBR), 5/7 (Non-GBR) |
| Wrong way driving warning [11] | 0.1 | 100 | $\leq 0.1$ | 8 | 1/3 (GBR), 5/7 (Non-GBR) |

users utilize the cell by receiving chunks of $3\,\mathrm{MB}$ in intervals of $1\,\mathrm{s}$.

## V. RESULTS AND DISCUSSION

Subsequently, the evaluated performance results for PayDA and further common scheduling procedures achieved by laboratory-based measurements and extensive simulations are presented and discussed. Further details regarding the implementation of PayDA and relevant aspects of the simulation environment are given in [1].

### A. Scenario 1: Homogeneous Data Traffic

Fig. 3 shows a comparison of measured and simulatively evaluated HOL delays for one RT and seven NRT users in the homogeneous data traffic scenario. The HOL delay refers to the time since the first data packet of a user's transmission buffer is queued. In the given worst-case scenario, the entire cell capacity for downlink transmissions is utilized by a total of seven users (one RT and six NRTs) resulting in one unscheduled UE in each scheduling iteration, which further increases the accumulated HOL delays. In case of RT users, the results show that the deadlines for time-critical data transmissions are complied due to less remaining time to the respective deadline. In contrast, this prioritization penalizes time-tolerant data transmissions leading to higher DMRs for NRT users. If a deadline is missed, the eNodeB schedules the data according to the best-effort principle and the corresponding receiving application has to decide if and how to process it. In Fig. 4, the overall average HOL delay of RT and NRT users validated by laboratory-based measurements and simulations is shown for both PayDA and EDF. For comparison, also the HOL delays resulting from the usage of other scheduling schemes are illustrated. As the number of users increases, the mean HOL delay also increases due to a packet flooding of the queues caused by the high cell load. The similar results of the laboratory-based measurements and the simulative evaluations prove the high suitability of the latter one enabling reliable performance analyses also for data traffic scenarios with much more users. Moreover, similar HOL delays can be achieved by using PayDA (simulative evaluation) and EDF, because the data traffic only differs in terms of the maximum delay margin. Finally, the results prove that PayDA is able to reduce the average HOL delay significantly compared to the ones caused by using Maximum Throughput (MT), Proportional Fair (PF) and Round Robin (RR).

### B. Scenario 2: Heterogeneous Data Traffic

In Fig. 5, the laboratory-based and simulatively evaluated average DMRs caused by using PayDA and EDF in the heterogeneous data traffic scenario for up to a total of eight users (one RT and up to seven NRT users) are illustrated. In addition, the resulting mean DMRs when using MT, PF and RR are shown. The results achieved by experimental measurements and by simulations prove that PayDA enables considerably lower average DMRs compared to the other scheduling mechanisms. In this respect, the absolute DMRs are quite high because the results also cover a very high cell utilization caused by up to seven NRT users leading to unscheduled users. Since all UEs have similar channel conditions due to the cabled setup, MT allocates available RBs equally to all users. Especially in high-load situations, this leads to a higher DMR in comparison to the other scheduling strategies. Moreover, PF and RR do not take deadlines into account at all so that the DMRs are higher compared to the ones when using PayDA and EDF. Finally, PayDA outperforms EDF in terms of lower average DMRs because data transmissions with similar deadlines but less remaining data are prioritized.

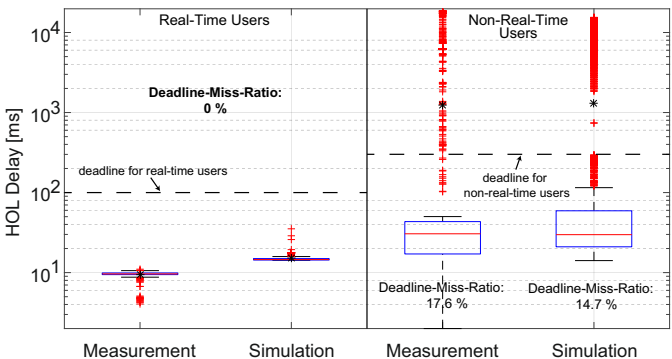

Figure 3. The high similarity of the HOL delays evaluated by measurements and simulations using PayDA in a worst-case scenario with homogeneous data traffic proves the high significance of the simulation results.

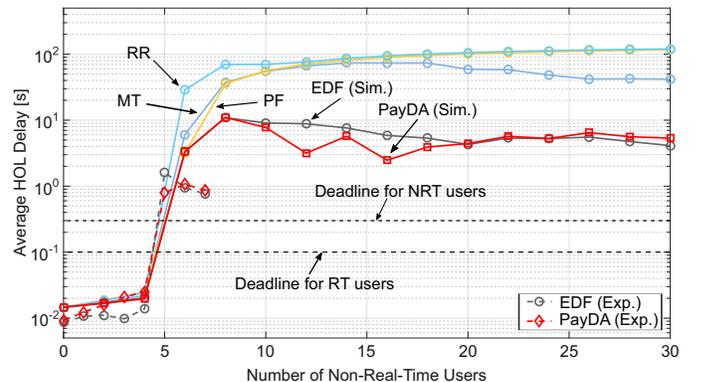

Figure 4. Average HOL delay of RT and NRT users for different scheduling schemes. The performance of PayDA is validated for a number of up to 8 users by laboratory-based measurements.

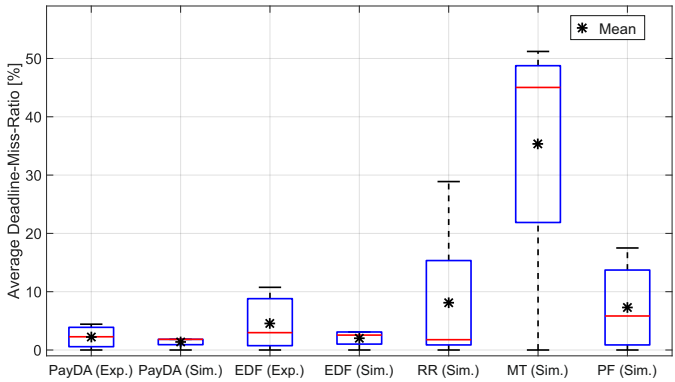

Figure 5. Average DMRs for different scheduling mechanisms in a heterogeneous data traffic scenario proving the high performance of PayDA regarding RT data transmissions validated by experimental measurements.

## VI. Conclusion

In this paper, we validated the performance of PayDA for different types of downlink data traffic with the help of measurements in a close-to-reality experimental laboratory setup and by means of simulation. In this regard, the PayDA scheduling scheme has been implemented for an SDR-based eNodeB to validate the performance results achieved by simulations. The results prove the high agreement of both the SDR-based and the simulation-based implementation of PayDA. While the experimental validation can be used for a small number of users due to the limitations of a laboratory setup, the simulations also allow for performance evaluations in large-scale data traffic scenarios. Compared to EDF, PayDA enables a resource-efficient and RT-capable packet scheduling by additionally considering the remaining packet sizes of each user's data stream in the scheduling metric and therefore is able to outperform EDF for the examined scenarios in terms of lower DMRs. With regard to homogeneous data traffic scenarios, PayDA always performs at least as well as EDF and for heterogeneous data traffic scenarios even improves the average DMR significantly. This makes PayDA suitable for improving the reliability of MTC and V2X RT applications in current and upcoming cellular networks. In the future, a numerical analysis would be of great interest to theoretically validate the performance of PayDA. The presented dual-stage validation setup consisting of both an SDR-based implementation in an experimental laboratory setup and extensive simulations can be used to enhance PayDA. Concerning this point, further performance analyses of PayDA in more complex scenarios including varying channel conditions and user mobility are of particular interest. Also, the effects of congestion control mechanisms used in wide-spread transmission protocols like Transmission Control Protocol (TCP) on the resulting user experience and possible interdependencies are in the focus of future research.

## Acknowledgment

This work has been supported by the Franco-German *BERCOM* Project (FKZ: 13N13741) co-funded by the German Federal Ministry of Education and Research (BMBF), the German Research Foundation (DFG) within the Collaborative Research Center SFB 876 "Providing Information by Resource-Constrained Analysis", project B4 and by the federal state of North Rhine-Westphalia and the "European Regional Development Fund" (EFRE) 2014-2020 in the course of the "CPS.HUB/NRW" project under grant number EFRE-0400008.